\documentclass[conference]{IEEEtran}
\IEEEoverridecommandlockouts
\usepackage{cite}
\usepackage{amsmath,amssymb,amsfonts}
\usepackage{algorithmic}
\usepackage[linesnumbered,ruled,vlined]{algorithm2e}
\usepackage{graphicx}
\usepackage{textcomp}
\usepackage{subfigure}
\usepackage{xcolor}
\usepackage{multirow}
\usepackage{chngcntr}
\usepackage{enumerate}
\usepackage[shortlabels]{enumitem}
\counterwithout{figure}{section}
\counterwithout{figure}{subsection}

\newtheorem{definition}{Definition}

\newtheorem{lemma}{Lemma}

\usepackage{color}

\title{Social Network of Extreme Tweeters: A Case Study\thanks{This works is partly funded by NSF Grant 1738411}}

\def\BibTeX{{\rm B\kern-.05em{\sc i\kern-.025em b}\kern-.08em
    T\kern-.1667em\lower.7ex\hbox{E}\kern-.125emX}}
\begin{document}
 
% author names and affiliations
% use a multiple column layout for up to three different
% affiliations
\author{\IEEEauthorblockN{Xiuwen Zheng}
\IEEEauthorblockA{San Diego Supercomputer Center\\
University of California San Diego\\
La Jolla, CA 92093\\
Email: xiz675@eng.ucsd.edu}
\and
\IEEEauthorblockN{Amarnath Gupta}
\IEEEauthorblockA{San Diego Supercomputer Center\\
University of California San Diego\\
La Jolla, CA 92093\\
Email: a1gupta@ucsd.edu}
}
\maketitle

\begin{abstract}
The number of posts made by a single user account on a social media platform Twitter in any given time interval is usually quite low. However, there is a subset of users whose volume of posts is much higher than the median. In this paper, we investigate the content diversity and the social neighborhood of these extreme users and others. We define a metric called ``interest narrowness'', and identify that a subset of extreme users, termed anomalous users, write posts with very low topic diversity, including posts with no text content. Using a few interaction patterns we show that anomalous groups have the strongest within-group interactions, compared to their interaction with others. Further, they exhibit different information sharing behaviors with other anomalous users compared to non-anomalous extreme tweeters. 
\end{abstract}


\section{Introduction}
\label{sec:intro}
Social Media is part of our daily lives, and increasingly more people are actively participating in various social media platforms. It was recently reported \cite{WParticle} that Twitter has 126 million daily users, with an estimated annual growth rate of about 9\%. It is estimated that roughly 46\% of Twitter users are on the platform daily. In this paper, we investigate the following questions: \textit{What types of users tweet an enormous amount and what do they talk about?}

The intuition behind this paper comes from the observation that we can characterize users' tweeting behavior based on the volume and the content diversity of their tweets.

\begin{figure}[t]
        \centering
\subfigure[]
{        \includegraphics[width=0.32\textwidth]{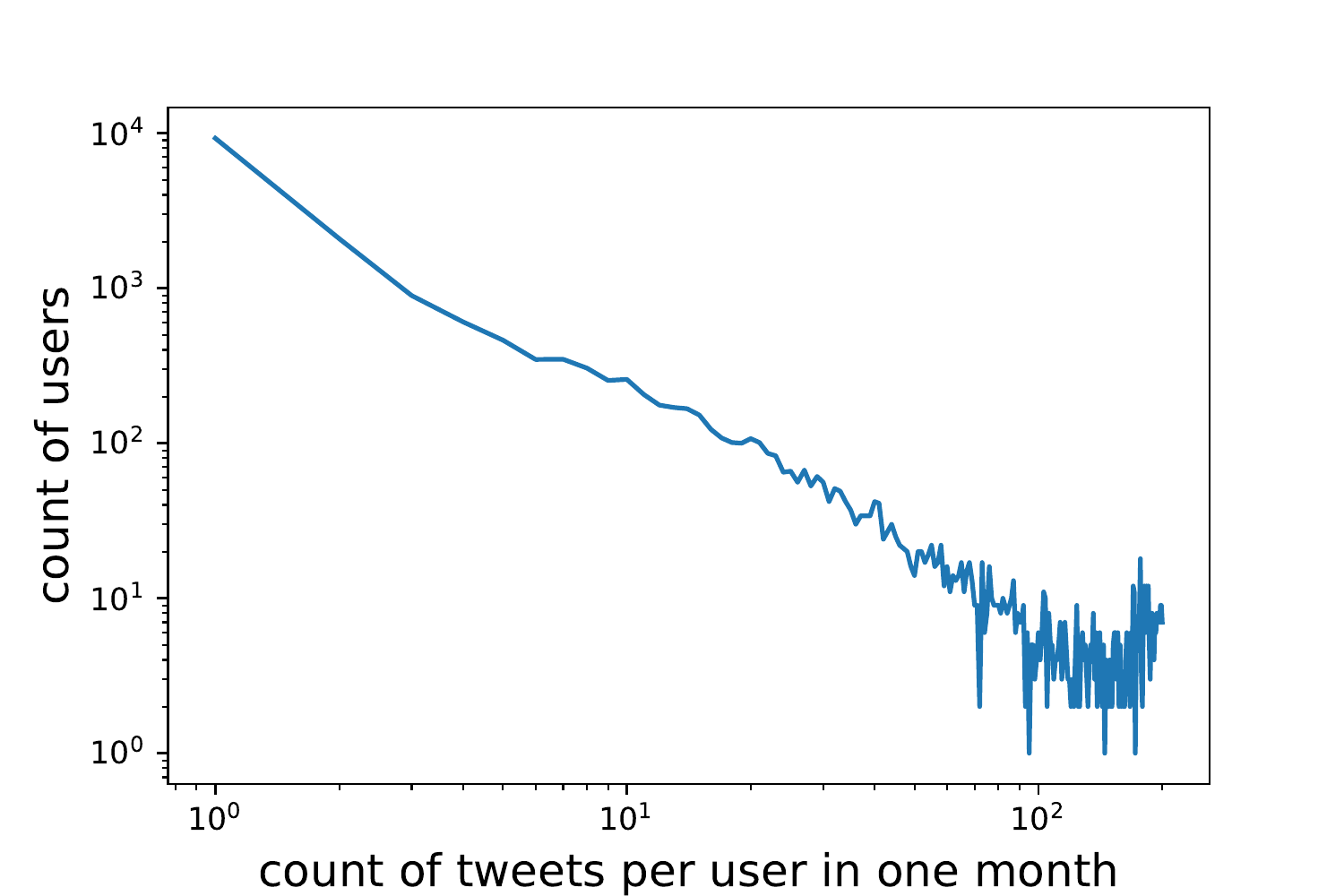}
\label{fig:tweetcounts}
}
\vspace{-1em}
\subfigure[]{
    \includegraphics[width=0.3\textwidth]{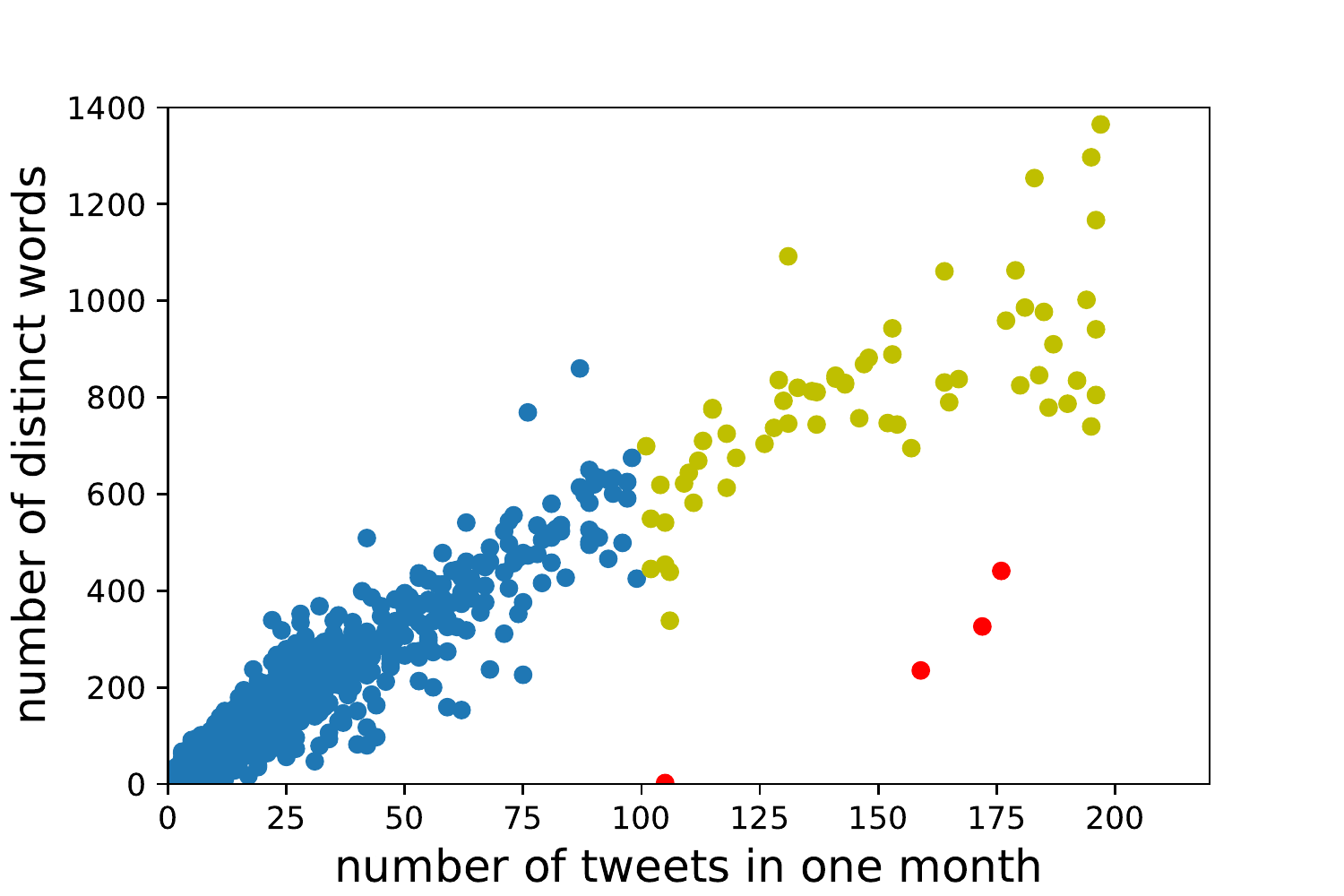}
\label{fig:wordcounts}
}
\caption{(a) Frequency distribution of tweet count for users in a month. (b) Frequency distribution of distinct words vs. number of tweets in a month.}
        \label{fig:sample_subfigures}
\vspace{-1em}
\end{figure}

We first consider tweet volumes for individual users. How often do people tweet? Based on our data set of over 1.5 billion tweets, we observe that over any arbitrary time interval, the number of tweets by a user follows a power law type of distribution (see Fig. \ref{fig:tweetcounts} for a typical distribution for one month) -- most users post very few tweets while only a few users write considerably more tweets. In the empirical result shown in Fig. \ref{fig:tweetcounts}, only 20\% of the users posted more than 24 tweets in July 2017. We use the term \textit{extreme tweeters} (ETT) for users who tweet more frequently than an average user in any given time interval (we give a more precise definition later). 

A different stratification of users can be created based on the \textit{content diversity} (c-diversity) of their posts. Intuitively, a user who is interested in many topics will have a higher content diversity than a user with a  narrow range of interests (e.g., only football). One simple, but rough measure of  c-diversity is the number of distinct words (not counting mentions) used by a user over all their posts in a given period of time. Fig. \ref{fig:wordcounts} shows a typical plot of the c-diversity of users as a function of the number of tweets they sent. In general, users who tweet more tend to have higher c-diversity. Clustering the frequency distribution reveals three different clusters corresponding to (a) users who tweet less and and use fewer distinct words (blue cluster), (b) users who tweet more and have higher content diversity (greenish yellow cluster), and (c) the small number of users who tweet more and yet have low content diversity (red cluster). 

In this case study, we explore the tweeting behavior as well as the social network of ETT, who constitute groups (b) and (c) above. We are particularly interested in group (c) because at first glance, their high-tweet-rate, low-diversity behavior is anomalous and appears somewhat counterintuitive. In the light of this exploration, the paper makes the following contributions:
\begin{enumerate}
    \item We propose a novel way to classify users based on their tweet rate and a new measure of verbosity called \textit{interest narrowness}.
    \item We present an algorithm to detect the anomalous users.
    \item We investigate the nature of the network relations of the anomalous group and the other groups.
    \item We show that the social interaction of anomalous and non-anomalous groups vary with time, and the vigorousness of the interaction intensifies around events like elections.
\end{enumerate}

\subsection{Related Work}
\label{sec:related}
The problem of user characterization in a social network has been investigated by many research groups. We present a few samples of these research efforts. 

A 2016 survey \cite{tuna2016user} covers a wide bandwidth of different approaches for user characterization. The behavioral properties they report range from conscientiousness and extroversion to privacy behavior, deceptive traits and response to social attacks. In contrast to our approach, many of the reported analyses studied in this paper are based on focused user surveys. 

Gabrowicz et al \cite{grabowicz2013distinguishing} take a bond-theory based approach and distinguish between social user groups and topical user groups in a network based on features like reciprocity, topicality and activity. Of these, topicality, defined based on a metric called ``normalized entropy'' measures how much the topics of discussion vary within a group. The higher the entropy, the greater is the variety of terms and, according to the theory, the more social the group is. However, this measure considers the words to be independent, which usually does not hold.

Similar to our notion of anomalous users, \cite{jang2013discovering} investigates the concept of ``dedicators'', users who transmit information in selected topic areas to the people in their egocentric networks. The concept of dedication is determined by volume, engagement, personal tendencies and topic weight, where personal tendency includes the user's topic diversity as measured by Latent Dirichlet Allocation (LDA), and engagement measures the activity level of conversations. 

Diversity of topics within text posts is also analyzed in papers like \cite{o2013characterizing} that perform LDA to compute topics and then determine topic diversity  across a user’s posts as the number of distinct probable topics found across all of the user’s posts. 

Closer to our application, Bail et al \cite{bail2016combining} used a POS-tagged BOW model to analyze Facebook posts for political content and derived a network of correlated concepts. They demonstrated how some advocacy organizations produce social media messages that inspire far-ranging conversation among social media users. Interestingly, their network analysis is based on the co-occurrence of concept terms from which they extract interesting connection patterns that characterize influence modalities for advocacy groups.

In contrast to all related work, ours is the first attempt, to our knowledge, that specifically analyzes the posting behavior and interaction patterns of extreme tweeters. Although we use Twitter as our example social media platform, our method is equally valid for any other platform that exhibits high-volume postings and vigorous interactions.
\section{The Setting}\label{sec:background}
%In this section, we first briefly introduce the Twitter data set used in this paper and conduct statistical study to capture the common traits of ``anomalous user'' \todo{change name}. We then put forward several metrics to quantify the abnormality for a user. Based on these metrics, the problem of detecting these abnormal users in social network is formulated, and a stream-based computation framework is also devised to make the detection task efficient and can be easily scaled to large-scale network. 

% \subsection{data set Description}
% \label{subsec:data}
% The primary data set for this case study is a collection of ``politics related'' tweets obtained by using the Twitter streaming API, using a set of keywords as a filter. A large majority of the collected data of 1.6 billion tweets spans from 2016 till date. The filtering keywords are names of politicians (members of the US Congress, members of the President's cabinet) and topics at the center of public debates (\texttt{healthcare, tax reform, Russia, abortion}$\ldots$). This collection is supplemented by a second set of tweets of all users who are the most active during the same period. For this case study, we have used a subset of our full collection, covering 6 weeks of data from October 1, 2018 till November 15, 2018, which is around the  US midterm election. The rationale for selecting this time period is that total number of tweets is much higher compared to other time intervals.

\subsection{User Behavior Classification}\label{subsec:user_study}
We now formalize the intuitive user behavior classification scheme presented in Section \ref{sec:intro}.
\subsubsection{Classification based on Tweet Volume}
Assume that $W$ is the time window of observation (e.g., 24 months), and $\Delta \ll W$ is the minimal analytic interval (MAI), i.e., a minimum time-interval (e.g., 1 week) over which data is collected in order to perform any user behavior analysis. We denote consecutive MAIs as $\Delta_1, \Delta_2 \ldots \Delta_n$ within $W$. %Let $P_{\Delta_i}^u(t)$ represent the probability distribution function that user $u$ has posted $t$ tweets in $i$-th MAI. 
\begin{definition}[ETT Behavior]\label{def:ETT}
Let  $U$ be the set of users, and $T^{u}(\Delta_i)$ be the number of  posts by a user $u\in U$ in $i$-th MAI  $\Delta_i$, then we say an  user \textbf{$u'$ exhibits ETT behavior in $\Delta_i$} if $T^{u'}(\Delta_i) \geq \mathbf{E}\left[T^u_{\Delta_i}\right] + \delta \cdot \sqrt{\mathbf{Var}\left[T^u_{\Delta_i}\right]}$, where $\delta \geq 0$ is an arbitrary constant to control the selectivity. \end{definition}

\begin{definition}[ETT Interval]
If a user $u$ exhibits ETT behavior in MAI \{$\Delta^u_i, \Delta^u_j \ldots \Delta^u_l$\}, then the \textbf{ETT interval of user $u$} is defined as the concatenation of consecutive subintervals from \{$\Delta^u_k$\}.
\end{definition}

Often a single user would not have a single continuous period of hyperactivity, but several ETT intervals within an observation window. We can use both the longest ETT interval (LETTI), and the total ETT interval (TETTI) as measures of a user's sustained ETT behavior. In this paper, we simply classify users with no associated ETT intervals as \textit{regular users}, and users with at least one ETT interval as \textit{ETT users}. 

% The 
% Fig.~\ref{fig:tweetcounts} shows a flat tail, and the number of users  decreases exponentially as tweet count increases. We also count the total number of tweets contributed by users with each tweets count during this month. Users with tweets count more than 24 contributes to over 90\% of total tweets on twitter and there are only around 20\% users who have such high twitter counts from Fig.~\ref{fig:tweetcounts}. 

\subsubsection{Classification based on Content Diversity}
In Section \ref{sec:intro}, we used the number of distinct words as a rough measure of c-diversity for a user. However, it is not a suitable measure of c-diversity because it actually measures the vocabulary diversity of a user. While a low-vocabulary user will have lower c-diversity than a high-vocabulary user, the measure cannot well distinguish between the c-diversities of two users with comparable vocabulary sizes. Secondly, just the raw count of words does not capture the thematic diversity of a user. Although two users have comparable vocabulary sizes, one may cover more themes than the other. In Section~\ref{subsec:topic_measure}, we develop a new measure called \textit{Interest Narrowness} by taking a topic model type approach.  

\section{Measuring Interest Narrowness}
\label{subsec:topic_measure}
We use the bag-of-words and singular value decomposition (SVD) techniques to develop the \textit{Interest Narrowness} measure. For a user $u$, we construct its text matrix $\mathbf{M}^u$ by adopting the bag-of-words model on all tweets of $u$ during a certain period of time. Let $p$ be the tweet count of $u$ and $q$ be the number of distinct words (except stop words and URLs) over all tweets of $u$. Apparently,  $\mathbf{M}^u$ is of dimension $p\times q$, and  without loss of generality, we assume that $p\leq q$. Apply SVD to $\mathbf{M}^u$:
\begin{equation}
    \mathbf{M}^u = \mathbf{U}\mathbf{\Sigma}^u \mathbf{V},
\end{equation}
where $\mathbf{U}$ and $\mathbf{V}$ are  $p\times p$  and $q\times q$ unitary matrices respectively and $\mathbf{\Sigma}^u$ is a  matrix where the diagonal entries $\sigma_{i}^u=\mathbf{\Sigma}^u_{i,i}$ are singular values.  Equivalently, $\mathbf{M}^u$ can be rewritten by a weighted sum of $p$ separable matrices: 
\begin{equation}\label{eq:svd}
    \mathbf{M}^u = \sum_i \mathbf{A}_i^u = \sum_i  \sigma_i^u \cdot \mathbf{U}_i \otimes \mathbf{V}_i,
\end{equation}
where $\mathbf{U}_i$ and $\mathbf{V}_i$ are the $i^{th}$ columns of $\mathbf{U}$ and $\mathbf{V}$ respectively, and $\otimes$ refers to the outer product. Based on  Eq.~\eqref{eq:svd}, we define the contribution of $j^{th}$ separable matrix $\mathbf{A}^u_j$ as follows,
\begin{equation}
    c^u_j = \frac{{\sigma_j^{u}}^2}{\sum_i \sigma_i^{u2}}.
\end{equation}

\subsubsection{Exact SVD Based Measure (EM)}
Assume that  separable matrices $\mathbf{A}^u_1, \cdots, \mathbf{A}^u_p$ are sorted in descending order by their corresponding singular values. Given a threshold $d\in [0, 1]$, let $K$ denote the minimum value of $k$ such that 
$$\sum_{j=1}^{k}c_j^u = \frac{\sum_{j = 1}^{k}\sigma_j^{u2}}{\sum_i \sigma_i^{u2}}\geq d.$$ 
Thus $K$ is the minimum number of $k$ such that the top-$k$ separable matrices can explain $d\times 100\%$ of matrix $\mathbf{M}^u$. In other words, the first  $K$ columns of $\mathbf{V}$ can represent most of the  topics of $u$'s tweets. Naturally, the interest narrowness of user $u$ can be defined as,

\begin{equation}\label{narrow}
   \gamma^u = 1 - \frac{K}{p}.
\end{equation}

Notably, the bottleneck of the exact measurement is the computation of SVD on text matrix $\mathbf{M}^u$, which takes time $O(\min(p^2q, pq^2 ))$.

\subsubsection{Randomized SVD based Measure (RM)}
Evaluation for the first measure for all ETT users is expensive  since both  the size of the  text matrix and the number of  ETT users can be large. To speed up the computation, we can approximate matrix decomposition  by using  the  randomized SVD \cite{halko2011finding} where  only partial singular values are computed. For a user $u$ with text matrix $\mathbf{M}^u$ of dimension $p\times q$, it can be approximated  by
\begin{equation}
    \mathbf{M}^u \approx \widetilde{\mathbf{U}}\widetilde{\mathbf{\Sigma}}^u\widetilde{\mathbf{V}},
\end{equation}
where $\widetilde{\mathbf{U}}$ is of $p\times k$, $\widetilde{\mathbf{V}}$ is of $k\times q$ and $\widetilde{\mathbf{\Sigma}}^u$ is of $k\times k$ where $k < \min(p, q)$. The interest narrowness is then given by,
% In our implementation, $k$ is set to $\max(10, p/10)$ and evaluate the topic narrowness by measuring how many percentage of original text matrix $\mathbf{M}^u$ been contributed by $10\%$ topics of its tweets count:
\begin{equation}\label{narrowness2}
    \eta^u = \frac{\sum_{j = 1}^ {k}\widetilde{\sigma_j}^{u2}}{\sum_i \sigma_i^{u2}} = \frac{\sum_{j = 1}^ {k}\widetilde{\sigma_j}^{u2}}{||\mathbf{M}^u||_F^2},
\end{equation}
where ${||\mathbf{M}^u||_F}$ is the Frobenius norm of matrix $\mathbf{M}^u$ and $||\mathbf{M}^u||_F^2 = \sum_{i = 1}^p\sum_{j=1}^{q} |m_{ij}^u|^2 = \sum_{i = 1}\sigma_i^{u2}$.  The commonly used implementation of randomized SVD takes time $O(pq \log k + (p + q)k^2)$ \cite{candes2009exact}.
Notice that $k$ serves as a hyper-parameter in the computation of narrowness. In our experiments we have noticed that setting $k$ to $\max(10, p/10)$ can well represent the c-diversity of all tweets. For this fixed setting of $k$, a large $\eta$ stands for relatively narrow topic interests.

Given any time interval $I = [t_s, t_e]$, we can define \textbf{\textit{anomalous users}} (shown as red cluster in Fig. \ref{fig:wordcounts}) as a user with  ETT behavior and  narrow topic interests during $I$.  Algorithm~\ref{algo:overview} provides the framework to detect anomalous users based on the definition.
% \begin{definition}[anomalous user]
% Given a time period denote as $T$, a suspect  user has two properties, 
% \begin{itemize}
%     \item \textbf{Big volume of tweets:} the user has a much larger number of tweets than normal users during $T$.
%     \item \textbf{Narrow topic interests:} The user has a much more narrow topic interests than random  tweets on twitter during $T$. 
% \end{itemize}

% \end{definition}

\begin{algorithm}
\footnotesize
\caption{\emph{Anomalous User Detection}}\label{algo:overview}
\KwIn{a list of triples: $\mathcal{T} = \{(T.u, T.text, T.time)\}$, a fixed time period: $I$, and  hyper-parameters:  $\delta$ and  $\lambda$}
\KwOut{a list of anomalous users: $\mathcal{A}$}
\tcc{Step 1: Find ETTs}
$T'\gets \{(T.u, T.text)| T\in \mathcal{T}, T.time \in I \}$\;
group by  users to get per user tweet count and tweet corpus  as  $\mathcal{C} \gets \{(C.u, C.count, C.all\_texts)\}$\;
 $ETT \gets \{C.u| C \in \mathcal{C}, C.count \geq \mathbf{E}\left[\mathcal{C}.count\right] + \delta \sqrt{\mathbf{Var}\left[\mathcal{C}.count)\right]}$\;
\tcc{Step 2: Calculate Narrowness}
 $N \gets max(\{C.count|C\in \mathcal{C}\})$\;
 get the maximum number of distict words for individual user $D \gets max(\{C.distinct\_word| C\in \mathcal{C}\})$\;
$\mathcal{H} \gets \{\}$\;
\If{$N \times D  \leq  M$}
{\For{$u\in ETT$}
{use EM (Eq.~\eqref{narrow}) to calculate  interest narrowness $nrw$ for $u$\;
$\mathcal{H}$.add($(u, nrw)$)\;}
}
\Else{
\For{$u\in ETT$}
{use RM (Eq.~\eqref{narrowness2})  to calculate  interest narrowness $nrw$ for $u$\;
$\mathcal{H}$.add($(u, nrw)$)\;}
}
\tcc{Step 3: Find Anomalous Users}
$\mathcal{A}\gets \{H.u| H \in \mathcal{H}, H.nrw \geq \mathbf{E}\left[\mathcal{H}.nrw\right] + \lambda \sqrt{\mathbf{Var}\left[\mathcal{H}.nrw)\right]}$\;

\end{algorithm}

Algorithm~\ref{algo:overview} takes a list of $(user, text, time)$ triples as input and outputs all users with ETT behavior and high interest narrowness.
The algorithm has two hyper-parameters,  $\delta$ and $\lambda$, which jointly control the selectivity of anomalous users.  Intuitively, the larger  $\delta$ and $\lambda$ are, the more strict the criteria of anomalous users will be. 
The algorithm consists of  two steps. Given a time period $I$, lines 1-3 constitute step 1 where all users who exhibit ETT behavior in $I$ are found based on Definition~\ref{def:ETT}. For each of these users, lines 4-14 calculate its interest narrowness. Notably, we set a threshold $M$ to decide whether or not  to apply the randomized SVD based measure (RM).  If the possible largest size of text matrix, i.e., $N\cdot D$, is smaller than $M$, the exact SVD based measure (EM) is adopted; otherwise, RM is performed for efficiency consideration. In our setting, $M$ is set to be $2,000 \times 5,000$, which is experimentally demonstrated to reach a balance between efficiency and effectiveness. Finally, line 15 finds out users from ETT with  high interest narrowness values as anomalous users.

The complexity of Algorithm~\ref{algo:overview} comes primarily from the computation of SVD for each user in $ETT$. If the exact version of SVD is adopted, the total time complexity is $O(|ETT|\cdot \min(ND^2, DN^2))$, where $N$ and $D$ represent the maximum tweet count and the maximum number of distinct words, respectively. Otherwise, the total complexity should be $O(|ETT|\cdot [ND \log k + (N + D)k^2])$.

We choose the randomized SVD based algorithm rather than the more standard LDA technique \cite{jang2013discovering,o2013characterizing} based on the following considerations. First, the number of potential topics is hard to set, especially in our setting where the topics are computed per user and we have potentially many users. Second, when the tweet count of one user is not big enough, SVD gives more meaningful results than LDA because the quality of LDA-produced topics is satisfactory when the training set is large. Finally, the  interest narrowness not only depends on the number of topics, but on within-topic term-diversity as well. 
\section{Social Network Analysis}
\label{sec:soc_network}
\subsection{Rationale}
\label{sec:rationale}
%The ETT behavior and anomalous user are defined only based on users' individual behavior,  and in this section, by introducing social network information during the same time period $T$, we try to understand their roles and behaviors in the networking. 
So far, we have focused on the ETT behavior of users and identified narrow-interest users with ETT behavior. We now explore the social network around these users to study their interaction patterns with other users, as well as the conversation topics of their network neighborhoods.
We would like to investigate four questions about anomalous users:
\begin{enumerate}[start=1,label={(\bfseries Q\arabic*):}]
    \item Do anomalous users  interact heavily with each other?
    \item Do anomalous users interact heavily with extreme tweeters  or regular users?
    \item If some anomalous users interact heavily with each other, what is their behavior pattern as a group?
    \item Do anomalous users  have different behaviors during different time periods, especially around major public-opinion-inciting events?
\end{enumerate}

The anomalous users themselves constitute a very small part of the larger social network that is induced by various forms of communication (reply, retweets, mentions, etc.) between any pair of users. We seek to identify connection patterns in network neighborhoods of the users of interest. In the following section, we present a few simple connection patterns amongst the anomalous users, the non-anomalous users with ETT behavior and regular users, and  Section \ref{sec:experiment} shows that these patterns suffice to bring out some distinctive characteristics of anomalous users.

\subsection{Three Simple Patterns}
\label{sec:patterns}
\begin{figure}[tbp]
    \centering
    \includegraphics[scale=0.45]{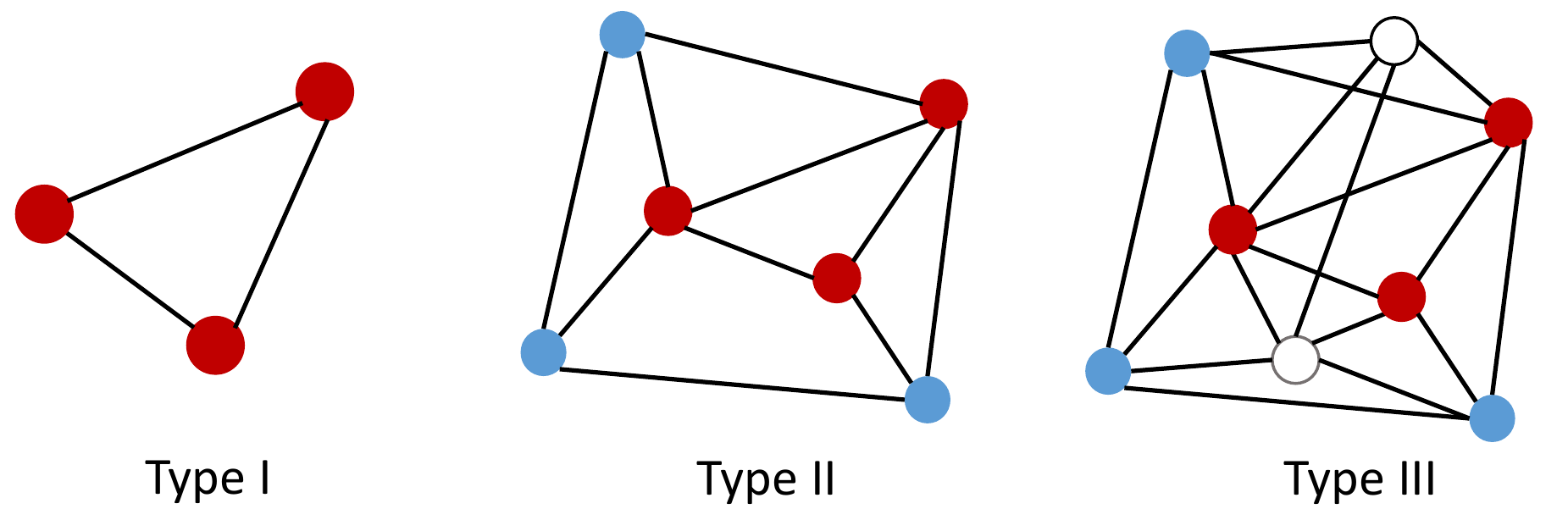}
    \caption{Three connection patterns around anomalous users.}
    \label{fig:social_network}
    \vspace{-1em}
\end{figure}

In Fig.~\ref{fig:social_network} we show three simple connection patterns involving the three user categories. There are three types of nodes in the pattern graph: red nodes represent anomalous users, blue nodes represent extreme tweeters  and white nodes represent regular users. Edges of the graph are undirected because in this case study we are simply exploring the connections and not the direction of messages that flow between user groups. 

The first connection pattern (called Type-I hereafter), denoted as  $G_1 = (V_1, E_1)$, shows only within-group triads for anomalous users. The edges represent the \textit{mentions} relationship between two anomalous users: $(v_i, v_j) \in E_1$ if $v_i$ mentions $v_j$ or  $v_j$ mentions $v_i$ during a certain time period $T$.  
The second connection pattern (Type-II), denoted as $G_2 = (V_2,  E_2)$, extends the triads to include first neighbors of anomalous users who are also extreme tweeters in the same time period. $V_2$ is the set of anomalous users and extreme tweeters who are mentioned by or mention an anomalous user, and edges $E_2$ also include the mentions between these non-anomalous ETT users. 
As for  the third case, the pattern network (Type-III) is formed by considering  anomalous users and \textit{all} their first neighbors. These three patterns correspond to the questions Q1 and Q2 posed in Section \ref{sec:rationale}.

%To figure out that if these anomalous users satisfy the second %property of anomalous group definition, i.e. densely connected, the %first graph   is built and anomalous users $\{S\}$ which form a %$k_1$-core where $k_1$ is the  largest coreness in the graph are %selected out. To figure out if these anomalous users have similar %target users, the second and third networks are built to find out %the $k_2$-cores and $k_3$-cores respectively consisting of  the set %of  anomalous users $\{S\}$ and some other users. $k_2$ and $k_3$ are 
%the largest coreness of  users in $S_1$ in the second and third %graph respectively. 
%Then the target similarity rate is calculated based on %Equ.~\ref{target_rate} to figure out if anomalous users in $\{S\}$ %have many common targets. At last, if the group $S$  satisfies the %first three properties, we measure the group topic interest %narrowness by combining all texts by each group member and %calculating narrowness by Equ.~\ref{narrow}. The narrowness is %compared with the average individual topic narrowness to figure out %if the group as a whole has more narrow interests than individuals.

\subsection{Metrics}
In this subsection, we propose three measures to assess the connectivity and density of our connection patterns in the network during any time interval. With these measures we expect to  evaluate the interaction intensity of anomalous users and their neighbors (Q1 and Q2), and capture the \textit{group behavior pattern} (Q3) of the anomalous user group and their immediate neighborhood.

\subsubsection{Coreness Distribution} $k$-core is a standard technique for tasks like dense region detection \cite{dorogovtsev2006k,alvarez2006large} and network feature extraction for different kinds of social network analysis \cite{ugander2012structural, huang2014cyber}. %We will use coreness distribution of all nodes in a graph to measure graph density. 
We recall that the coreness of a vertex in a graph is $k$ if it belongs to a $k$-core, but not to a $(k+1)$-core.     In this paper, we  use  the terms ``coreness'' and  ``core number'' interchangeably.

Let $G^T$ be the complete social network within time interval $T$. To capture the holistic connective characteristic of anomalous users and their related users in $G^T$, we introduce the concept of ``coreness distribution'', which is the distribution of coreness of nodes in the largest subgraph of $G^T$ satisfying a certain connection pattern (i.e., Type-I, Type-II or Type-III). Specifically, to measure the connection intensity among anomalous users, we extract the largest subgraph of $G^T$ satisfying the Type-I pattern, denoted by $G^T_1$, which would include all anomalous users, and then calculate core number of all users in $G^T_1$  to get the coreness distribution.  As $G^T_1$ only includes the  relations among anomalous users, the coreness distribution of $G^T_1$ would give an answer to Q1 from Section \ref{sec:rationale}. If the distribution shows that a subset of users have relatively high coreness, we can conclude that these anomalous users interact intensively with each other. Similarly, to capture how intensely do anomalous users interact with extreme tweeters (or regular users), the largest subgraph of $G^T$ satisfying  Type-II pattern (resp. Type-III pattern), say $G^T_2$ (or $G^T_3$), is constructed, and the coreness distribution of anomalous users in $G^T_2$ (or $G^T_3$) is calculated  and compared with that of $G^T_1$. 

\subsubsection{Measure of Collective Behavior}
The coreness distribution depicts the holistic interaction intensity of all anomalous users with different categories of users, which answers Q1 and Q2. However, it is possible that an   anomalous user  may not strongly connect with other anomalous users. We  focus on the group of strongly connected  anomalous users (Q3), defined as follows.
 
\begin{definition} [Anomalous Group]\label{def:group}
Let $G_A^T$ be a subgraph of $G^T$  satisfying,
\begin{itemize}
    \item \textit{Pattern Constraint:} $G_A^T$ satisfies Type-I pattern;
    \item \textit{Strongest Connection:} $G_A^T$ is a   $k$-core where $k$ is the largest degeneracy value among all subragphs of $G^T$ satisfying Type-I pattern;
    \item \textit{Maximality:} There does not exist another $k$-core satisfying Type-I pattern that is a supergraph of $G_A^T$. 
\end{itemize}
Then the nodes of $G_A^T$, say $\mathcal{A}$, form one anomalous group with coreness $k^\mathcal{A}_1 = k$. 
\end{definition}

\begin{lemma}\label{lem: maximality}
For any graph $G$ satisfying Type-$x$ connection pattern, $x = 1, 2, 3$, any subgraph  of $G$ also satisfies Type-$x$ pattern.
\end{lemma}

Based on Lemma~\ref{lem: maximality}, 
an anomalous group is, by definition, the maximal $k$-core of $G^T_1$ where $k$ is the maximum coreness value of nodes in $G^T_1$. 
In Type-II or Type-III connection patterns,  users in an anomalous group may act collaboratively (i.e., they    mention similar users), or they may have diverse behaviors (i.e., they interact with  different users). To measure such interaction patterns, we define the coreness of anomalous group and then  present two metrics.

\begin{definition} [Coreness of Anomalous Group]
 For an anomalous group $\mathcal{A}$, let $G^T_{A_x}$ be a subgraph of $G^T$ such that,
 \begin{itemize}
    \item $G^T_{A_x}$ satisfies Type-$x$ pattern where $x$ can be 1, 2, or 3.
     \item The nodes of $G^T_{A_x}$ is a superset of $\mathcal{A}$.
     \item $G^T_{A_x}$ is a $k^\mathcal{A}_x$-core where $k^\mathcal{A}_x$ is the largest degeneracy value among all subgraphs of $G^T$ satisfying Type-$x$ pattern and including nodes $\mathcal{A}$.
     
     %\item \textit{maximality:} {There does not exist another $k$-core satisfying Type-$x$ pattern and including $\mathcal{A}$ that is a superset of $G^T_{A_x}$.}
 \end{itemize}
 We say $k^\mathcal{A}_x$ is the coreness of $\mathcal{A}$ in Type-$x$ pattern.
\end{definition}

Obviously, $k^\mathcal{A}_2$  and $k^\mathcal{A}_3$  would be at least $k^\mathcal {A}_{1}$ because Type-II or III is an extension of Type-I pattern. From Lemma~\ref{lem: maximality}, $k^\mathcal{A}_2$  and $k^\mathcal{A}_3$ can be calculated  
 as $\min\{C = \{c_u | u\in \mathcal{A}\}\}$, where $c_u$ is the coreness of $u$ in $G^T_2$ or $G^T_3$, which are the largest subgraphs of $G^T$ satisfying Type-II or Type-III patterns respectively.

We extract the maximal $k^\mathcal {A}_{2}$-core and $k^\mathcal {A}_{3}$-core including $\mathcal{A}$ from $G^T_2$ and $G^T_3$ respectively. These cores could be considered as a group of users who highly interact with $\mathcal {A} $.  To evaluate  the behaviors of  group $\mathcal {A}$,  we provide the following measures. 
\begin{definition}[Common  Neighbor Ratio and Diversity Ratio]
 Let $\mathcal {A} $ be an anomalous group, and 
$\mathcal{U}$ denote the set of other nodes in the   $k^\mathcal {A}_{2}$-core or $k^\mathcal {A}_{3}$-core of  subgraph $G^T_2$ or $G^T_3$. Define the common  neighbor ratio (CNR) as,
\begin{equation}\label{target_rate}
   r = \frac{\sum_{u\in \mathcal{U}}{N_u}}{|\mathcal{A}||\mathcal{U}|},
 \end{equation}
where $N_u$ is the number of anomalous users in $\mathcal {A} $ who are connected to a user $u$. The diversity ratio (DR) is defined as,
\begin{equation}
    \beta = \frac{|\{u | u\in \mathcal{U}, N_u \geq r\times |\mathcal{A}|\}|}{|\mathcal{A}|}.
\end{equation}

 \end{definition}
 %$r$ can evaluate that, in average, for any user in the core,  how many percent of anomalous users in the group have interaction with it. 

\begin{figure}
    \centering
    \includegraphics[scale=0.48]{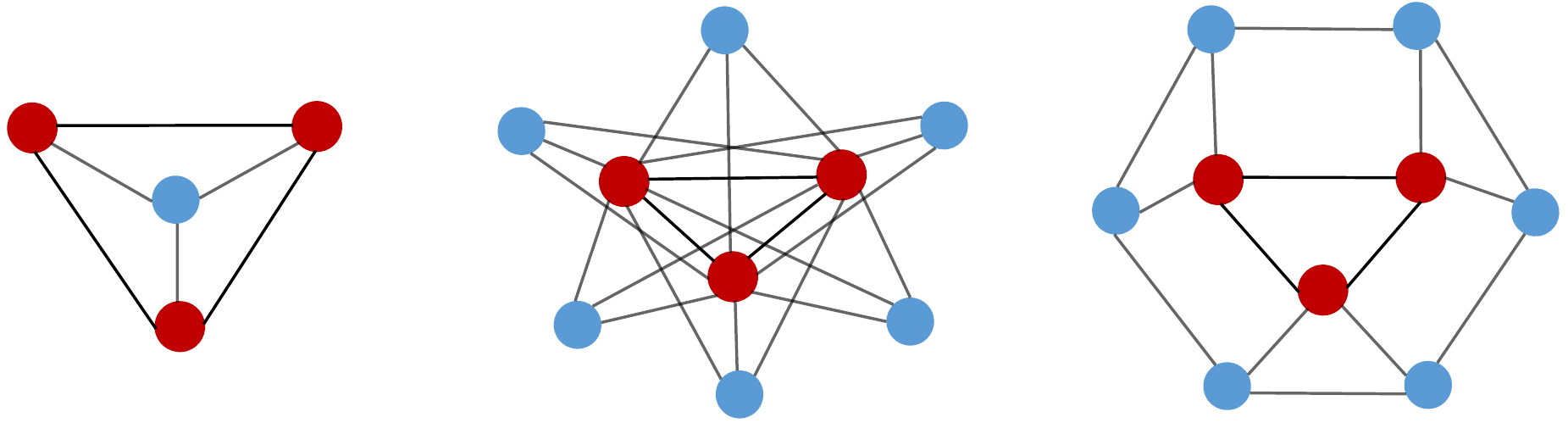}
    \caption{Illustration of group behavior patterns.}
    \label{fig:illustration}
    \vspace{-1em}
\end{figure}

 The common neighbor ratio $r$ can be interpreted as the fraction of anomalous users that can interact with a user in the core on an average. A large  $r$ implies that these anomalous  users  interact with others collaboratively in this core.  The diversity ratio $\beta$ can be interpreted as the fraction of users that have interaction with the anomalous group. Fig.~\ref{fig:illustration} illustrates three typical patterns. The first one shows a pattern with large $r$ and small $\beta$ ($r = 1$ and $\beta = 1/3$). 
 %$r = 1$ because the only blue  user in this core are attached to all anomalous users, and $\beta = \frac{1}{3}$.  
 For  groups with this pattern, group members, i.e., anomalous users,   have similar mention behavior and they only interact with a few users. The second pattern has large $r$ and large $\beta$ ($r = 1$ and $\beta = 2$). For groups with this pattern,  group members  have similar behavior, in addition,  they interact with a large number of users collaboratively.  For  the third pattern with small $r$ and large $\beta$  ($r = 1/3$ and $\beta = 2$), group members show diverse mention behaviors. 

%Hence, from the subgraphs that match the Type-I pattern, we select users in the $k$-core with the largest $k$ as a group $G$. Then in  Type-II network, we select the $k$-core with the largest $k$ that include all group members in $G$. Then the following two measures is to measure their behaviors as a group in the second core. There are different kinds of group behavior as the following shows.

\section{Experiment results}
\label{sec:experiment}
The primary data set for this case study is a collection of ``politics related'' tweets obtained by using the Twitter streaming API, using a set of keywords as a filter.  The filtering keywords are names of politicians (members of the US Congress, members of the President's cabinet) and topics at the center of public debates (\texttt{healthcare, tax reform, Russia, abortion}$\ldots$). A large majority of the collected data of 1.6 billion tweets spans from 2016 till date. The experiments conducted in this section is on a subset of the political tweet data set. We chose the subset to capture a time interval where the total number of messages were very high with many conversations, as described below.

%We selected a 6-week time period around the 2018 United States mid-term elections held on November 6, 2018. Our experiments are based on tweets from  October 1 to November 14, 2018 and break it down to 6 short time intervals  each of which has 7 or 8 days.  \comment{do you think we can move this part to extreme tweeter analysis? YES}
%Table.~\ref{tab:overview} describes simple statistics of  our data set during each time interval. 
%\begin{table}[htbp]\label{tab:overview}
%\caption{Overview of our data set}
% \scriptsize
% \begin{center}
% \begin{tabular}{|c|c|c|}
% \hline
% Time Period & Number of Users &  Maximum Tweet Count of a single user\\
% \hline
% Oct 1 - Oct 7 &   52,223 &  202\\\hline
% Oct 8 - Oct 15& 76,931 & 587\\\hline
% Oct 16 - Oct 23& 92,628 & 249\\\hline
% Oct 24 - Oct 31& 122,574 & 2,740\\\hline
% Nov 1 - Nov 7 & 96,855& 2,211\\\hline
% Nov 8 - Nov 14 &56,532&1,140 \\\hline
% \end{tabular}
% \end{center}
% \end{table}
% \comment{I think this table can be deleted because Fig~\ref{fig:ett_percent} % covers its information \ag{OK}}
% During  Oct 24 - Nov 7, there are much more active users than other time  % periods and users seem to tweet more within this time range. 

\subsection{Extreme Tweeters Analysis}
\begin{figure}[t]
    \centering
    \includegraphics[scale = 0.36]{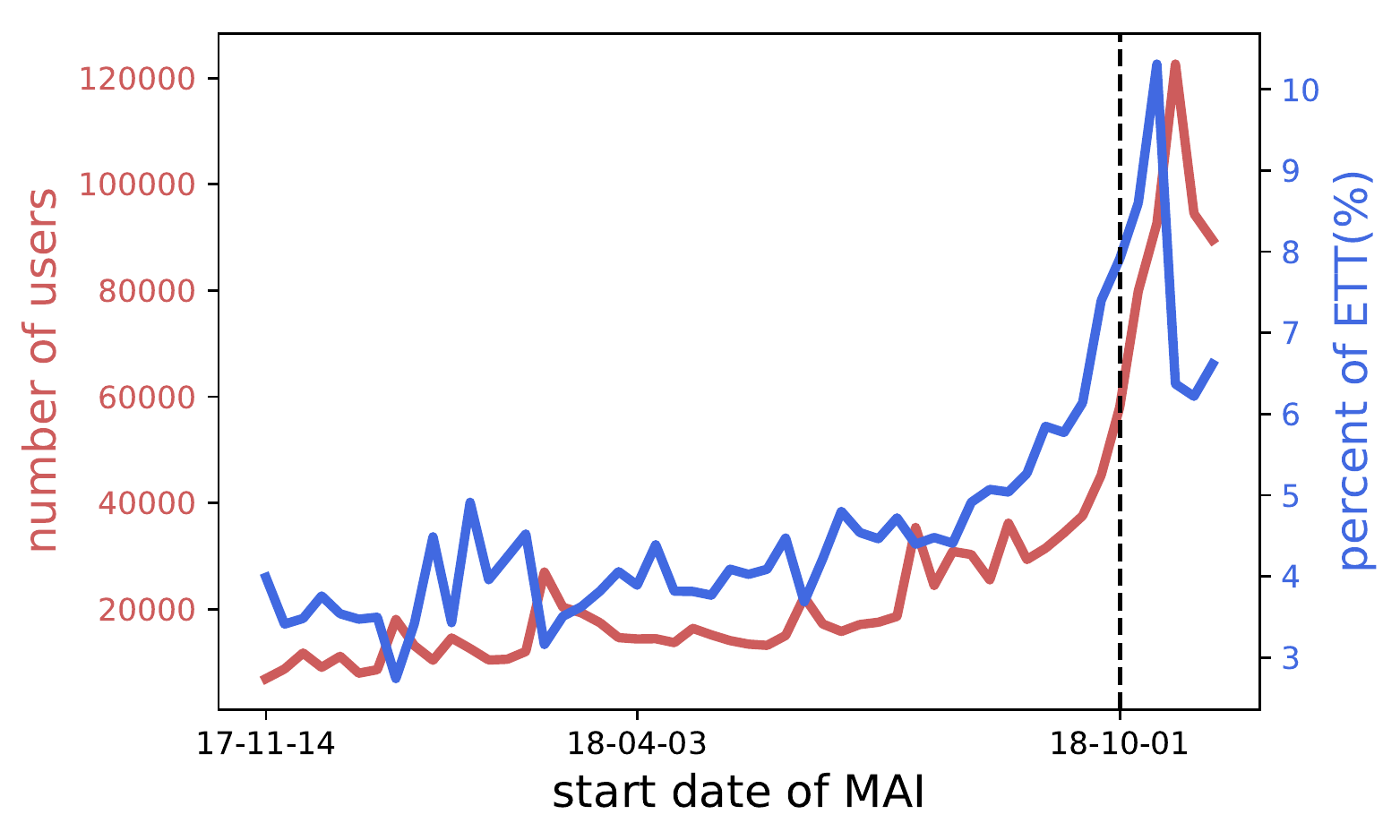}
    \caption{Percentage of users with ETT behavior at each MAI. The largest number exceeds 12000, over 10\% of the number of users.}
    \label{fig:ett_percent}
    \vspace{-1em}
\end{figure}

To analyze ETT behaviors, the total time window of observation $W$ is set as 12 months from Nov 14 2017 to Nov 14 2018 and the minimal analytic interval (MAI) $\Delta$ is set as one week. %There are totally 54 MAIs in our observation window. 
We set $\delta$ from Definition~\ref{def:ETT} as 1.5 and get all users with ETT behavior in each MAI illustrated in  Fig.~\ref{fig:ett_percent}. The red line shows the total number of users who tweet at least once during each MAI, and the blue line  shows the percentage of users with  ETT behavior.  Before September 2018, there are only a few users with ETT behaviors, however, after September  the fraction of ETT users  grows significantly, and reaches a peak at around the mid October of 2018. To get a reasonable number of ETT users for our experiments, we select a 6-week time period around the 2018 United States mid-term elections held on November 6, 2018. Specifically, our experiments are based on tweets from  October 1 to November 14, 2018, partitioned into 6 nearly equal time intervals.

\begin{figure}[t]
    \centering
    \includegraphics[scale = 0.38]{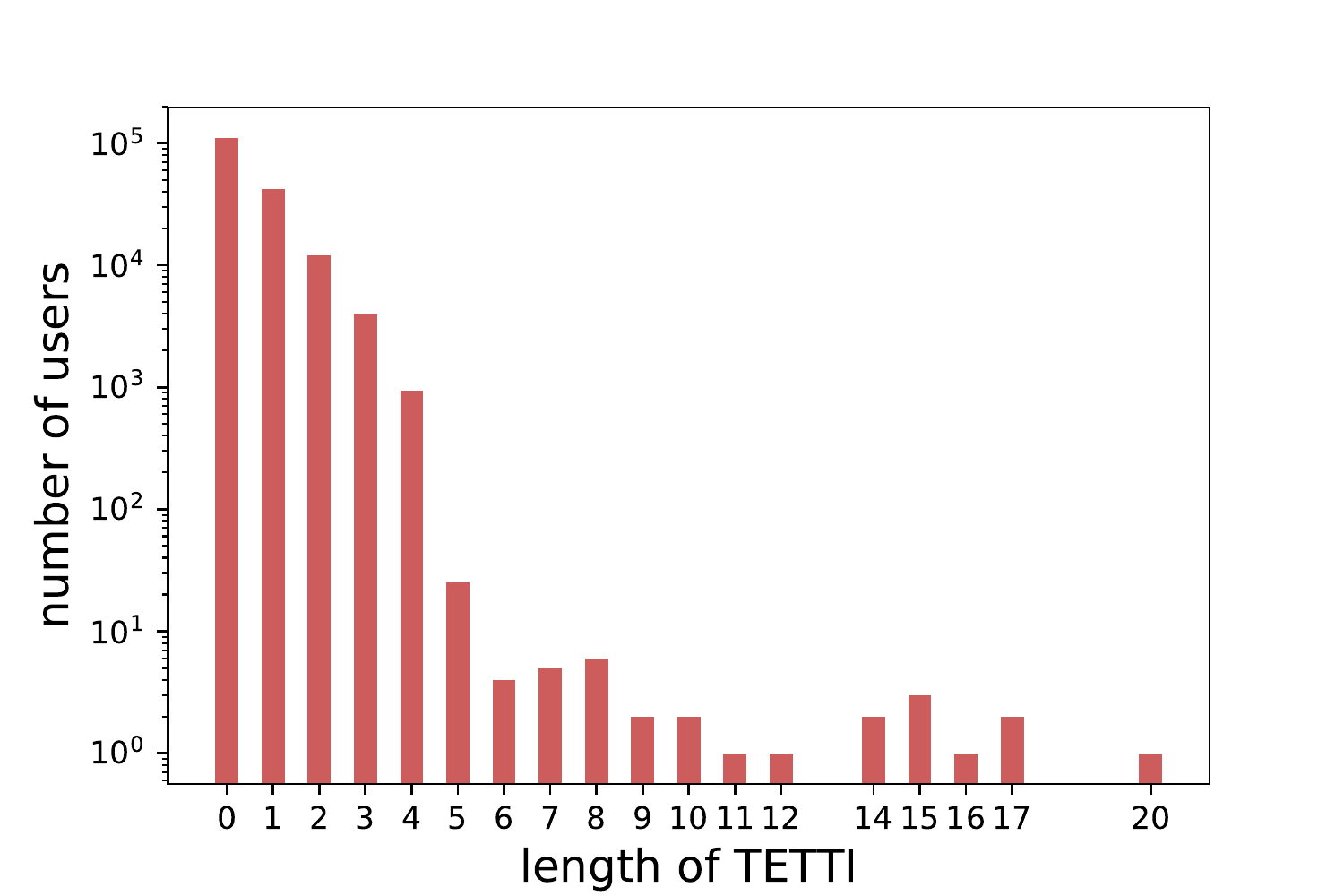}
    \caption{Distribution of users over TETTI length.}
    \label{fig:tetti}
    \vspace{-1em}
\end{figure}

The length of total ETT intervals (TETTI) for each user is calculated and Fig.~\ref{fig:tetti} shows the distribution of users over  TETTI lengths. Most users are regular users who never have ETT behavior. Only a few number of users have ETT behaviors frequently. The largest TETTI length within the 54 observation MAIs is 20. 

\subsection{Interest Narrowness Analysis}
Anomalous users are selected out for  each time period by following the framework illustrated in  Algorithm~\ref{algo:overview}.
The two hyper parameters, $\delta$ and $\lambda$, are set to  1.5 and 1 respectively. Table~\ref{tab:overview} shows the fraction of ETT users and the fraction of anomalous users (AU in Table~\ref{tab:overview}) over ETT users  during each time period. 

\begin{table}[htbp]
\caption{Percentages of ETT users and Anomalous Users}
 \scriptsize
 \begin{center}
 \vspace{-1em}
 \begin{tabular}{|c|c|c|c|}
 \hline
 Time Period & Users Count &   ETT (\% of Users)& AU (\% of ETT) \\
 \hline
 Oct 1 - 7 &   52,262 &  7.54 & 8.15\\\hline
 Oct 8 -  15& 76,935& 8.57& 8.51\\\hline
 Oct 16 -  23& 89,810 & 10.31& 9.38\\\hline
 Oct 24 - 31& 122,570& 6.14& 11.69\\\hline
 Nov 1 - 7 & 96,858& 7.36& 11.03\\\hline
 Nov 8 -  14 &69,060&8.19& 7.92\\\hline
 \end{tabular}
 \end{center}
 \label{tab:overview}
 \vspace{-1em}
 \end{table}
 
\subsubsection{Change of interest narrowness over different time periods}
\begin{figure}
\vspace{-1em}
    \centering
    \includegraphics[width = 0.34\textwidth]{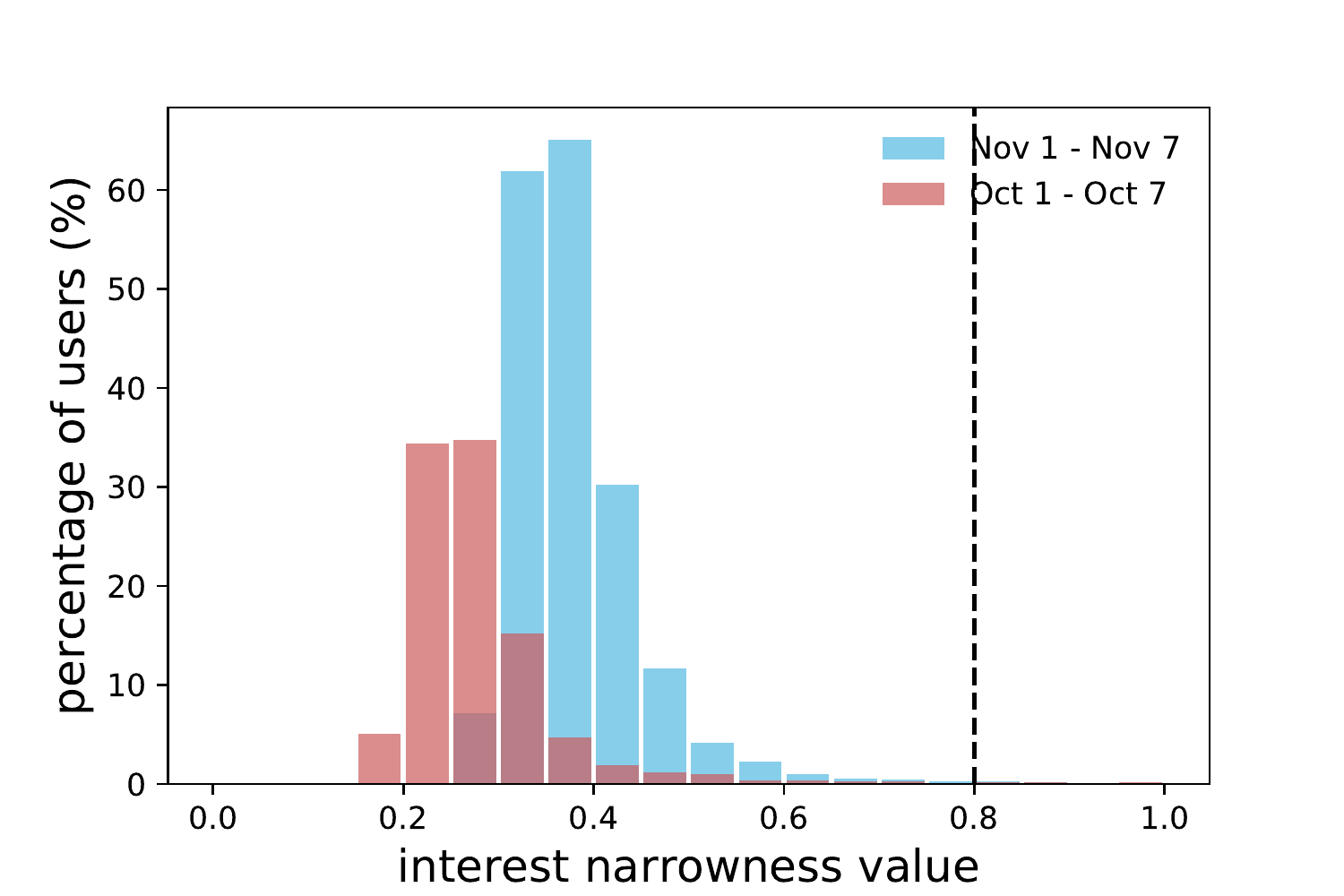}
    \caption{Extreme tweeters distribution over interest narrowness }
    \label{fig:narrowness}
    \vspace{-1em}
\end{figure}

 We calculate  the interest narrowness for every extreme tweeter at each time period. Fig.~\ref{fig:narrowness} shows the distribution of extreme tweeters over interest narrowness value at two time periods:  Oct 1 - 7 and  Nov 1 - 7.

The user distribution shifts to  right when time range changes from Oct 1 -  7 to Nov 1 - 7 and  more extreme tweeters concentrate on higher narrowness, i.e.,  extreme tweeters have narrower topic interests during Nov 1 - 7. 

\subsubsection{Analysis of users with extreme interest narrowness}
As the black dashed line in Fig.~\ref{fig:narrowness} shows, a few users have  extremely high interest narrowness, i.e., only $10\%$ of their tweets count number of  topics can well  represent all their tweets.  In order to understand their tweeting behaviors, for each  time period, we select users with interest narrowness  higher than $80\%$ and dive into their tweets. Many of these users have extremely high narrowness because they have a lot of ``null text tweets'' where they do not write any of their own words but only mention  others. We select out users with more than $80\%$ null text tweets and  Table~\ref{tab:rslt} shows their total tweets count and the number of users mentioned by them during a certain time period. 

\begin{table}[htbp]
\caption{Users with more than 80\% of null text tweets}
\scriptsize
\begin{center}
\begin{tabular}{|c|c|c|c|}
\hline
Time Period  & Users Count & Tweets Count &  Mentions Counts\\
\hline
Oct 1 - 7 &4 & 459 &  203 \\\hline
Oct 8 - 15& 9& 906 & 1,158 \\\hline
Oct 16 - 23&13 & 1,956& 1,284    \\\hline  
Oct 24 - 31&5& 2,459 & 1,058\\\hline
Nov 1 - 7 & 4  & 2,600 &  3,222\\\hline
Nov 8 - 14& 4 &  725 & 129\\\hline
\end{tabular}
\end{center}
\label{tab:rslt}
\end{table}

% \begin{table}[htbp]
% \caption{Some Users with Extreme Narrow Topic Interests}\label{tab:some_user}
% \scriptsize
% \begin{center}
% \begin{tabular}{|c|c|c|c|c|}
% \hline
% Time Period & User & Null Text Tweets Count&  Mentions count\\
% \hline
% Oct 8 - Oct 15 & A & 108 &  1,029\\\hline
% Oct 16 - Oct 23& B &  129 &  432 \\\hline
% Oct 24 - Oct 31& C& 819  & 756\\\hline
% Nov 1 - Nov 7 & C& 970  &  1,509\\\hline
% Nov 1 - Nov 7 & D& 612& 1,066\\\hline
% \end{tabular}
% \end{center}\label{tab:narrow}
% \end{table}

During Oct 8 - Nov 7,  a handful of users contribute many null text tweets and there are a large volume of users  mentioned by them. The situation becomes the most extreme during Nov 1 - 7 when   four users have 2600 tweets in total, of which more than 80\% have no self-written words,  and  mention more than 3000 distinct users. 
%\comment{maybe the following part can be deleted?}
% Table.~\ref{tab:narrow} shows  the mention statistics of some extreme users  and their node degrees (using the mentions % edge) during the time period. These users are significant because they send a significant number of tweets \textit{with % no textual content}.

% To observe their roles in social network, we sample a % set of  nodes including these extreme users from the % largest  subgraph satisfying the Type-II pattern and % plot the subgraphs for two time periods: from Oct 8 - % Oct 15 and from Nov 1 - Nov 7  as % Fig.~\ref{fig:special_user} shows. The red nodes stand % for anomalous users,  blue nodes stand for candidate % users and the red stars in the two figures  stands for % user A  and user C respectively from % Table.~\ref{tab:some_user}.  The two figures show that % the two users with high percentage of null text tweets % play very import role in social network  as they % interact with many other users. 

%  \begin{figure}[t]
%         \centering
% \subfigure[Oct 8 - Oct 15]
% {        \includegraphics[width=0.35\textwidth]{./figu% re/strange_user_social_1_10-8.pdf}
% }
% \subfigure[Nov 1 - Nov 7]{
%     \includegraphics[width=0.35\textwidth]{./figure/st% range_user_social_2.pdf}
% }
% \caption{Sample Subgraphs with extreme users.}
%         \label{fig:special_user}
% \end{figure}
%  

\subsection{User Distribution over Coreness}
%In order to analyze the interaction and connection of anomalous users on social network, we  built three types of social networks for each time period. 
To analyze the social connectivity between  anomalous users and other user categories, we construct the largest subgraphs from the complete social network that satisfying Type-I, Type-II and Type-III connection patterns for each time period. We call them Type-I, Type-II, Type-III social network  in the whole experiment section. 

\subsubsection{Type-I Social Network}
\begin{figure}[htbp]
    \centering
    \subfigure[Type I Network]{
    \begin{minipage}[t]{0.25\textwidth}
    \centering
    \includegraphics[width = \textwidth]{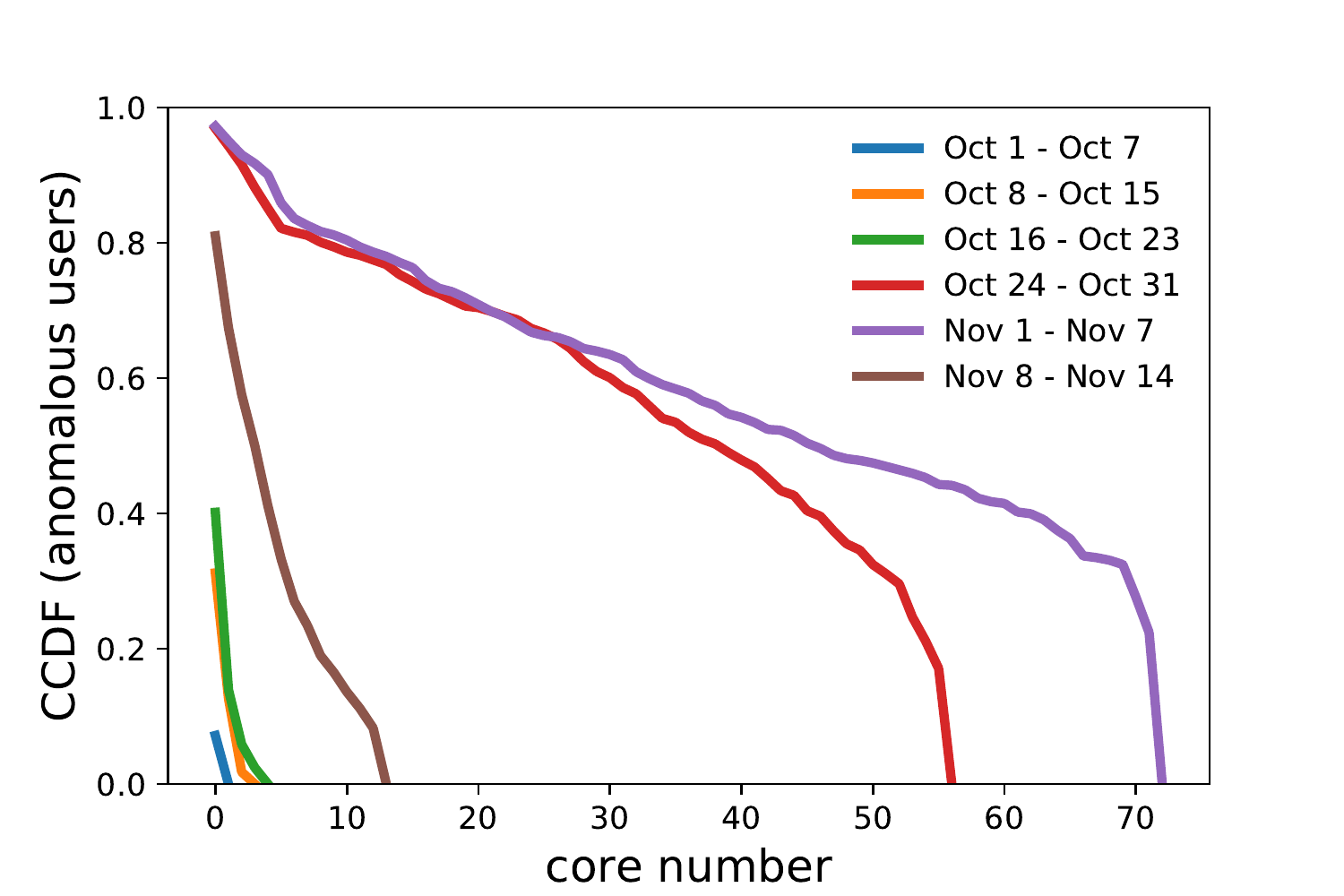}
    \label{fig:typeI}
    \end{minipage}%
    }%
    \subfigure[Type II Network]{
    \begin{minipage}[t]{0.25\textwidth}
    \centering
    \includegraphics[width=\textwidth]{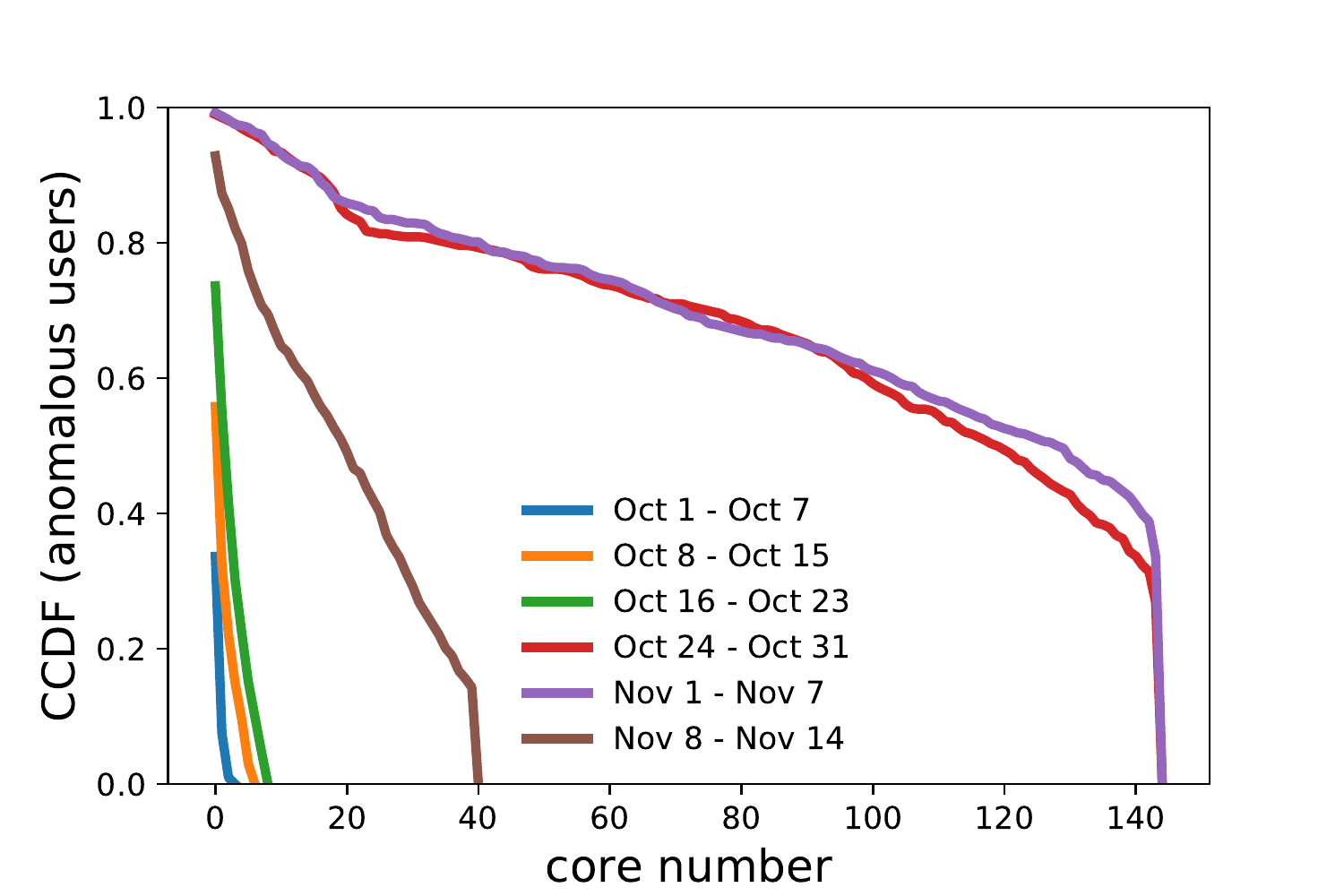}
    \label{fig:typeII}
    \end{minipage}%
    }%
%    \vspace{-1em}
    \centering
    \caption{CCDF of anomalous  users over coreness.}\label{fig:suspect_cdf1}
        \vspace{-1em}
\end{figure}

We evaluate the distribution of anomalous users over  coreness in Type-I social network at different time periods, and  Fig.~\ref{fig:typeI} shows the complementary cumulative distribution function (CCDF) of anomalous users. In the first three weeks of October, most users do not interact with each other and more than half of anomalous users have 0 coreness. However, when it is near the American midterm election, i.e., from Oct 24 - Nov 7,  the red and purple lines decrease  slowly at the beginning and  go down quickly at the tail, suggesting that users concentrate  on high coreness. There are only around 700 anomalous users at Nov 1 - 7 from Table~\ref{tab:overview}, while the largest coreness is  72, which implies extremely strong connections  amongst them.  However, after the election, the largest coreness dropped to  13, which means that many anomalous users  leave  hot interaction with others. In addition, the distribution of users  tends to be  more uniform  since the brown line decreases smoothly. 
 %users tend to have uniform distribution
 %, which means  that some people had  few interaction, while some were  interacting with others. 
\vspace{-1em}

% \begin{figure}[htbp]
%     \centering
%     % \subfigure[Oct 1 - Oct 7]{
%     % \begin{minipage}[t]{0.25\textwidth}
%     % \centering
%     % \includegraphics[width=\textwidth]{./figure/anomalous_dis0pdf.pdf}
%     % %\caption{fig1}
%     % \end{minipage}%
%     % }%
%     % \subfigure[Oct 8 - Oct 15]{
%     % \begin{minipage}[t]{0.25\textwidth}
%     % \centering
%     % \includegraphics[width=\textwidth]{./figure/anomalous_dis1pdf.pdf}
%     % %\caption{fig2}
%     % \end{minipage}%
%     % }%
%     % \vspace{-1em}
%      \vspace{-2em}
%     \subfigure[Oct 16 - Oct 23]{
%     \begin{minipage}[t]{0.25\textwidth}
%     \centering
%     \includegraphics[width=\textwidth]{./figure/anomalous_dis2pdf.pdf}
%     %\caption{fig2}
%     \end{minipage}%
%     }%
%     \subfigure[Oct 24 - Oct 31] {
%     \begin{minipage}[t]{0.25\textwidth}
%     \centering
%     \includegraphics[width=\textwidth]{./figure/anomalous_dis3pdf.pdf}
%     %\caption{fig1}
%     \end{minipage}%
%     }%
%          \vspace{-1em}
%          
%     \subfigure[Nov 1  to Nov 7]{
%     \begin{minipage}[t]{0.25\textwidth}
%     \centering
%     \includegraphics[width=\textwidth]{./figure/anomalous_dis4pdf.pdf}
%     %\caption{fig2}
%     \end{minipage}%
%     }%
%     \subfigure[Nov 8 - Nov 14]{
%     \begin{minipage}[t]{0.25\textwidth}
%     \centering
%     \includegraphics[width=\textwidth]{./figure/anomalous_dis5pdf}
%     %\caption{fig2}
%     \end{minipage}%
%     }%
%     \centering
%     \caption{Core Number Distribution of Type I  Network.}\label{fig:result}
% \end{figure}

\subsubsection{Type-II and III Social Networks}
\begin{figure}
\vspace{-1em}
    \centering
 \includegraphics[width=0.27\textwidth]{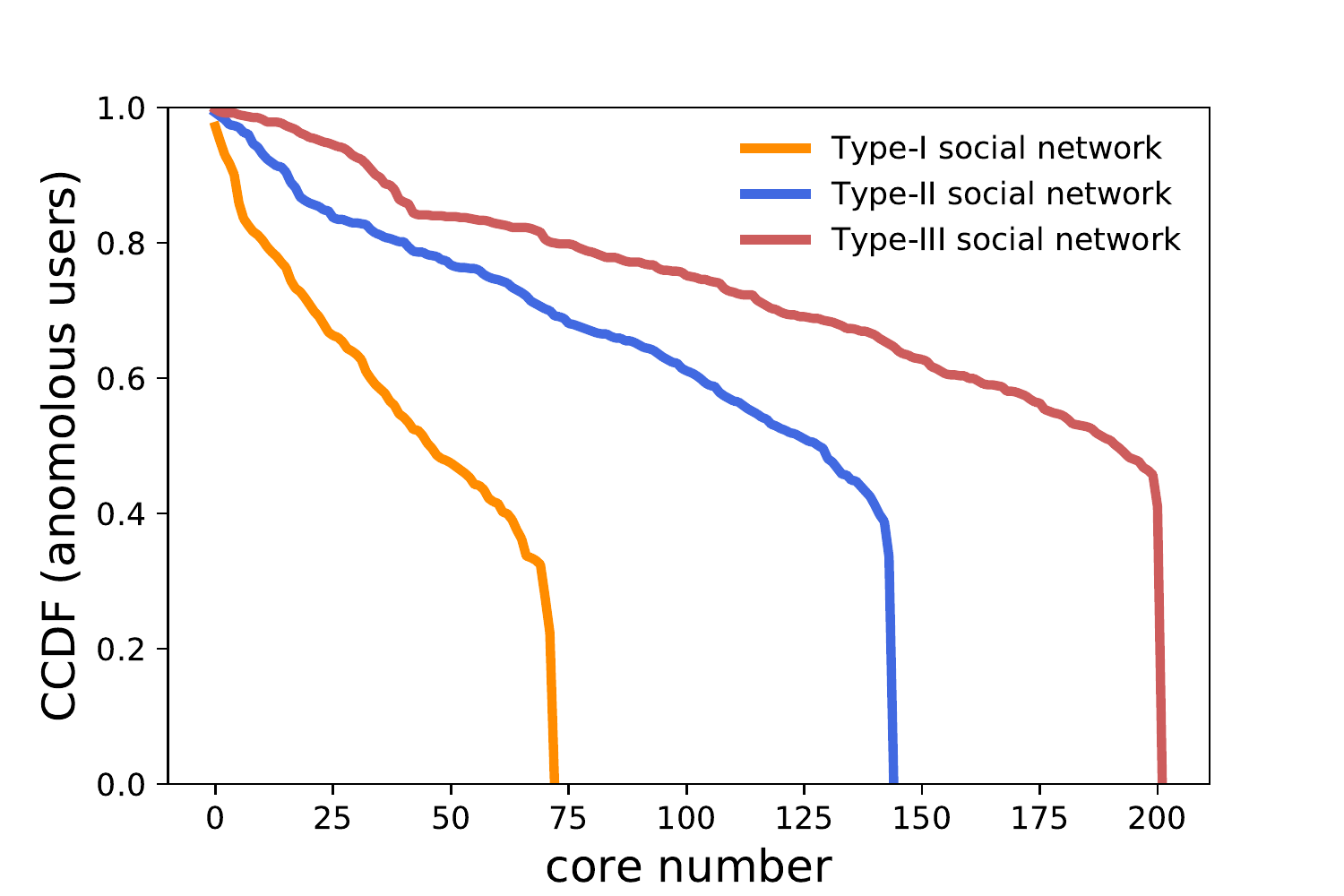}
    \caption{ CCDF of anomalous users in three networks during Nov 1 - 7}
    \label{fig:suspect_cdf2}
        \vspace{-2em}
\end{figure}

We evaluate  the coreness distribution of anomalous users when their ETT- or regular neighbors are included.  Fig.~\ref{fig:suspect_cdf2} shows the CCDF  of anomalous users over coreness in three types of networks during Nov 1 - 7. When including first neighbors,   the coreness of anomalous users increases dramatically, suggesting that many anomalous users have strong connections with their neighbors. The yellow line reaches  bottom before coreness 75, but there are around 70\% and 80\% of anomalous users  with coreness larger than 75 in  Type-II and Type-III networks respectively. In addition, the largest coreness  increases from  72 to around  140 and 200. However, from Table~\ref{tab:overview}, the number of anomalous users is only around 11\% of ETTs and less than 0.8\% of regular users at Nov 1 - 7. Considering the sizes of  these three  user categories, the increase of coreness of anomalous users is of less significance.
Thus, we conclude that many anomalous users indeed have strong connections with some extreme tweeters or regular users, however, from a holistic view of three  categories, interactions amongst anomalous users themselves are stronger than those between anomalous users with ETT users or regular users.

\subsubsection{Change of Social Network over Time}
To analyze the dynamic pattern of social networks, we also plot the CCDF of anomalous users for  Type-II social  networks during all observation  periods. As shown in  Fig.~\ref{fig:suspect_cdf1}, both  Type-I and Type-II  networks  show  the similar trend of changes. 
At the first three time periods, users  do  not involve in  dense interaction with each other  and the lines decrease rapidly at the beginning and reach  0 before 15 coreness. However, during Oct 24 - Nov 7, the lines suddenly and dramatically shift to  right  and they decrease slowly at first but  much faster at the tail, which means that many users interact intensively with each other.   However, after midterm election,  the lines (brown lines) shift to  left and  social networks tend to become more silent, but users still have more interactions than normal days, e.g., Oct 1 - 7. In a word, the interaction intensity between users changes over time, especially during important real world events.

\subsection{Anomalous Group Behavior Pattern}
For  first three weeks of our observation window, there is no anomalous group formed because anomalous users do not interact heavily during these time intervals.  We find an anomalous group in the other three time periods. Table~\ref{tab:group} presents their group  coreness of Type-I connection pattern (denoted as Coreness 1) and of Type-II pattern (denoted as Coreness 2), and  their common neighbor ratio and diversity ratio  in Type-II connection pattern.

\begin{table}[htbp]
\caption{Anomalous Groups}
\scriptsize
\begin{center}
\begin{tabular}{|c|c|c|c|c|}
\hline
Time Period & Coreness 1 & Coreness 2 &  CNR& DR\\
\hline
Oct 24 - 31& 56 & 143& 0.37& 1.59\\\hline
Nov 1 - 7 & 72 & 144 & 0.40& 0.89\\\hline
Nov 8 - 14& 13 & 37 &0.21&4.62 \\\hline
\end{tabular}
\end{center}
\label{tab:group}
\vspace{-1em}
\end{table}

\begin{figure}[t]
        \centering
\subfigure[Group I]
{        \includegraphics[width=0.26\textwidth]{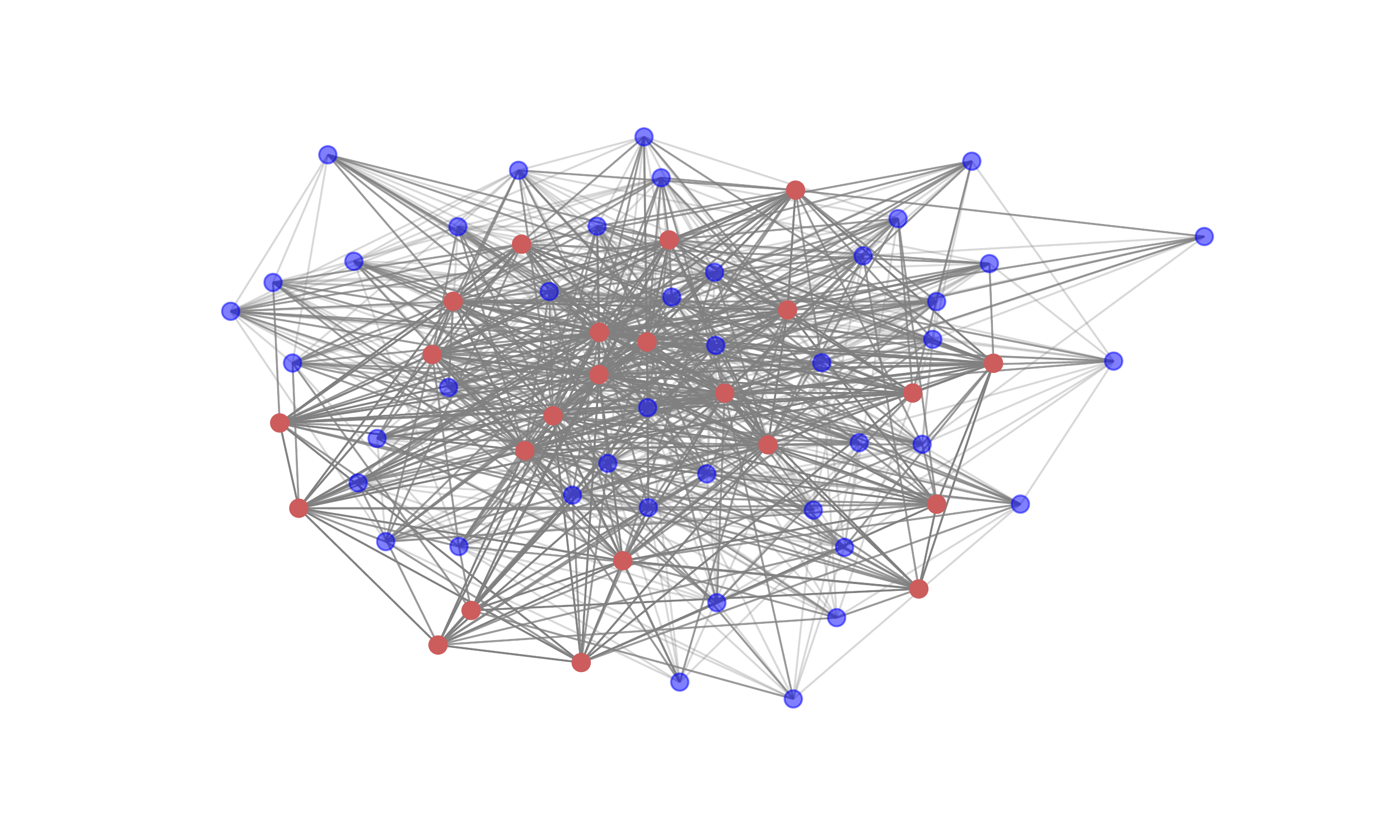}
}
\vspace{-1em}
\subfigure[Group II]{
    \includegraphics[width=0.26\textwidth]{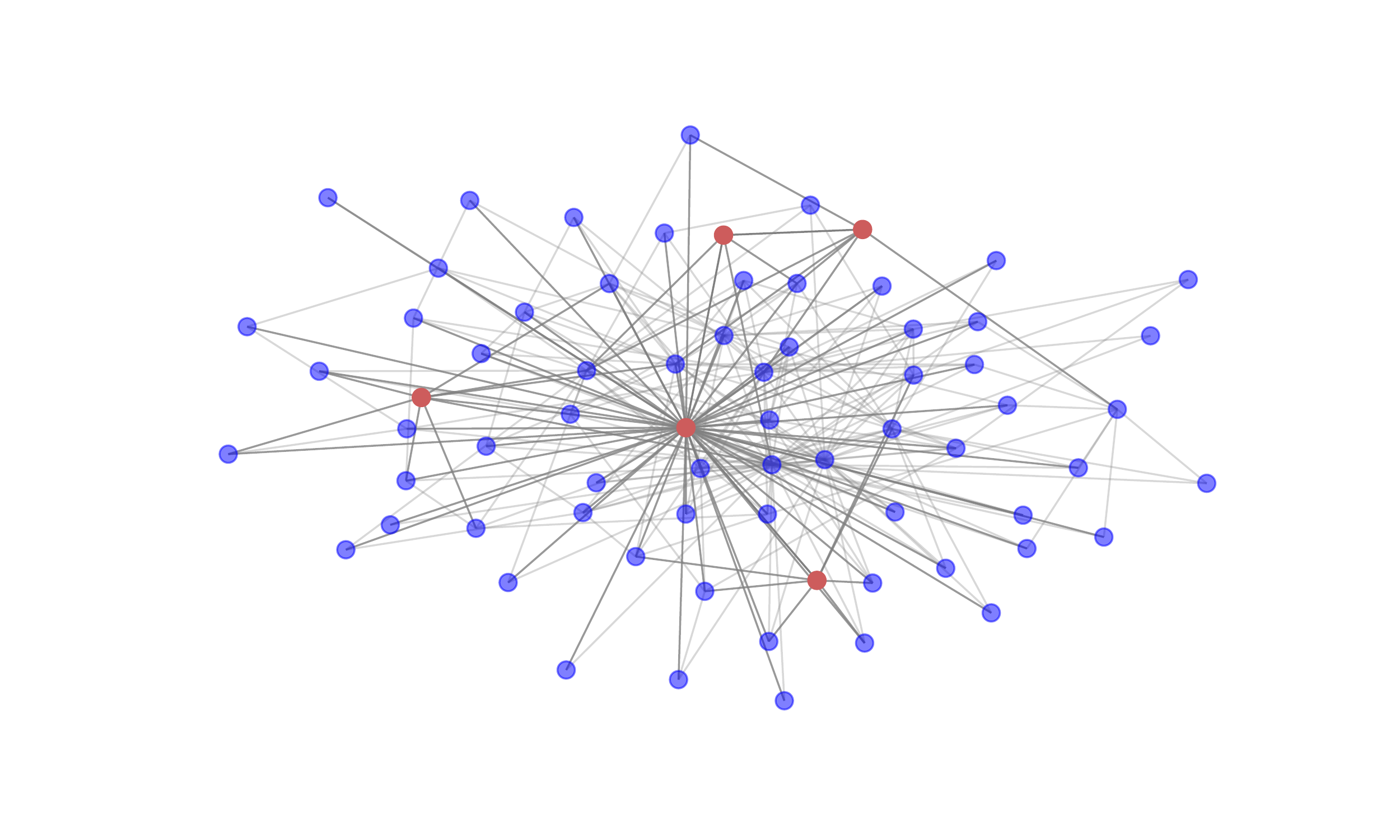}
}
\caption{Sample subgraphs satisfying  Type-II connection pattern.}
        \label{fig:pattern2}
    \vspace{-1em}
\end{figure}

 During Nov 1 - 7, as the relatively high CNR and low DR values suggest, the  anomalous users in this group (Group I) tend to have similar mention behaviors and they do not interact with many people comparing to their group size. Instead, during Nov 8 - 14, anomalous users in the group (Group II) tend to have diverse behaviors, i.e., they do not interact with others collaboratively and they mention a lot of people. Fig.~\ref{fig:pattern2}  shows two  sample subgraphs satisfying  Type-II connection pattern for these two groups. The red nodes denote users from the anomalous group and blue nodes denote extreme tweeters who connect with at least one anomalous user.

%\subsection{Topic Analysis for Anomalous Groups}

\begin{figure}[htbp]
    \centering
    \subfigure[Group I]{
    \begin{minipage}[t]{0.23\textwidth}
    \centering
    \includegraphics[width = \textwidth]{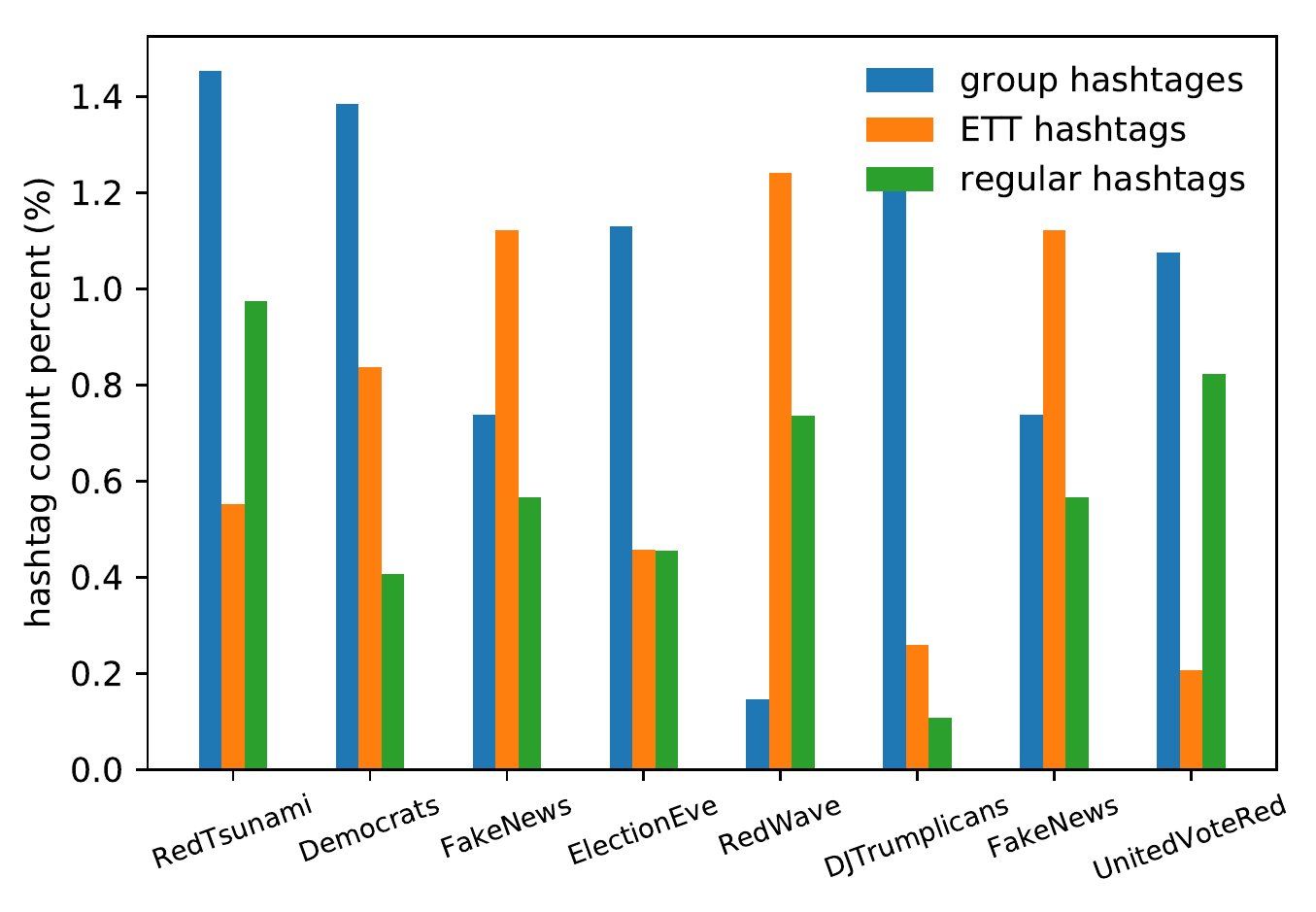}
    %\caption{fig1}
    \end{minipage}%
    }%
    \subfigure[Group II]{
    \begin{minipage}[t]{0.23\textwidth}
    \centering
    \includegraphics[width=\textwidth]{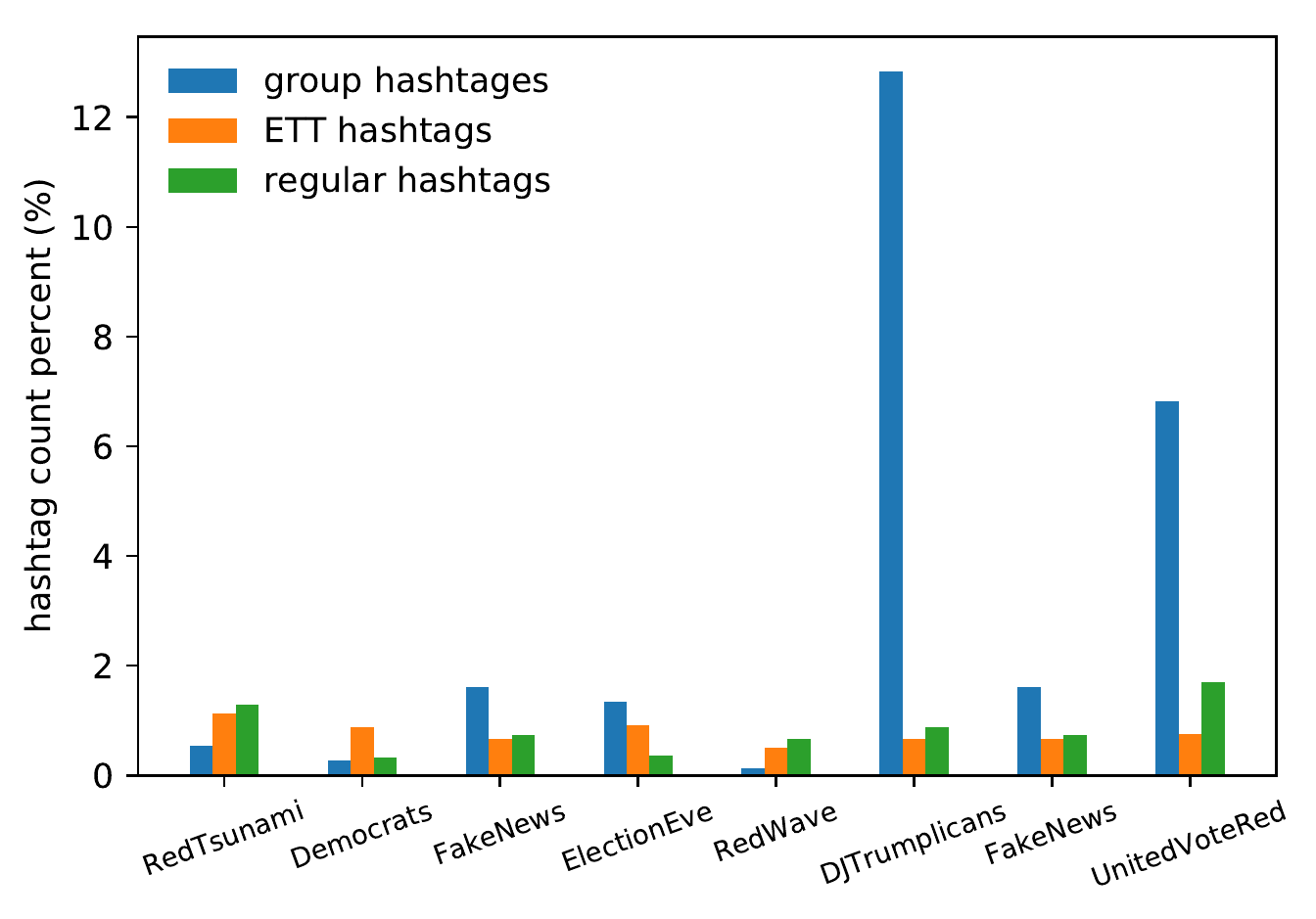}
    %\caption{fig2}
    \end{minipage}%
    }%
    \centering
    \caption{Hashtag histogram for anomalous group.}\label{fig:hashtag}
    \vspace{-1.5em}
\end{figure}

The two anomalous groups  show completely different mention behaviors. 
%To understand their conversation topics when they mention others, 
%Finally, we compare the conversation topics from these two anomalous groups.
%from different time intervals, Nov 1 - 7 and Nov 8 - 14 respectively. 
We collect all hashtags used by members of each group to mention i) regular users (regular hashtags), ii) non-anomalous ETT (ETT hashtags) and iii) each other within group (group hashtags). Fig.~\ref{fig:hashtag} shows the count percentages  of a same subset of hashtags  for the two  groups. %The blue, yellow and green bars  represents group hashtags, ETT hashtags and  regular hashtags respectively. %To understand whether they have different intra-group and inter-group hashtag uses, 
We calculate the correlation coefficient between counts of  group hashtags  and ETT hashtags  (\texttt{coef1}) and that between counts of group hashtags  and regular hashtags  (\texttt{coef2}) as shown in Table~\ref{tab:anomalous_group_stats}. Besides, the standard deviations (\texttt{stdev}) for group hashtags, ETT hashtags and regular hashtags distributions (\texttt{stdev1}, \texttt{stdev2}, \texttt{stdev3} respectively in Table~\ref{tab:anomalous_group_stats}) are calculated for each group. Note that the \texttt{stdev} presented in this section is standardized by  the percentage of hashtag counts.
\vspace{-1em}
\begin{table}[htbp]
\caption{Statistics of Hashtag Distributions}
\scriptsize
\begin{center}
\begin{tabular}{|c|c|c|c|c|c|}
\hline
  & \texttt{coef1} &  \texttt{coef2}& \texttt{stdev1} &\texttt{stdev2}& \texttt{stdev3}\\
\hline
Group I & 0.81  &0.77 & 0.004  & 0.003 & 0.002\\\hline
Group II  & 0.31& 0.10 & 0.024& 0.008  & 0.004\\\hline
\end{tabular}
\end{center}
\label{tab:anomalous_group_stats}
\vspace{-1em}
\end{table}

For Group I,  it has similar intra-group and inter-group hashtag usage behaviors as two correlation coefficients indicate, and all the three distributions have very low \texttt{stdev}, indicating  a uniform use of hashtags. For Group II, however, the low coefficients suggest that this  group may have  different hashtag uses when they mention amongst group members versus mention other categories.  In addition, the distribution of group hashtags  (blue bars in Fig.~\ref{fig:hashtag}) has much higher \texttt{stdev} than the other two distributions (yellow and green bars in Fig.~\ref{fig:hashtag}).  It indicates that some hashtags (e.g., DJTrumplicans or UnitedVoteRed) are used disproportionately often when members of  Group II interact amongst themselves.  However, when they mention other categories of users, their hashtag distribution is more uniform, i.e., the conversation covers broader topics.

During the first three time periods in our observation window, only one or two users in Group II are ETT users. However, since Oct 24,  most members of Group II show sustained hyperactivity and high  interest narrowness, i.e., they are  anomalous users during the other three time periods. Table~\ref{tab:anomalous_group} presents their group coreness in Type-I connection  pattern  and the \texttt{stdev} of  three hashtag distributions.  For each time period, this  group shows highly-skewed  hashtag usage within group, but broader interests outside it.
%\vspace{-1em}
\begin{table}[htbp]
\caption{Behaviors of Group II During Different Time Periods}
\vspace{-1em}
\scriptsize
\begin{center}
\begin{tabular}{|c|c|c|c|c|}
\hline
Time Period & Coreness &  \texttt{stdev1} &\texttt{stdev2}& \texttt{stdev3}\\
\hline
Oct 24 - 31& 54 &0.040 & 0.004& 0.005\\\hline
Nov 1 - 7 & 62& 0.015 & 0.003& 0.003\\\hline
\end{tabular}
\end{center}
\label{tab:anomalous_group}
\vspace{-2em}
\end{table}

\section{Conclusion}
\label{sec:conclusion}
%In this case study, we provide a platform-independent stratification mechanism for social media users based on their posting rates and topic narrowness. We define a novel metric to measure the  interest narrowness of users based on  post content and present an algorithm to classify users into three categories.  To analyze their interaction behaviors on social network, we define three simple connection patterns centered at anomalous users or extreme tweeters, and accordingly, we provide three metrics on these patterns to reveal the intra-group and inter-group behaviors of anomalous users. Our experiments suggest that our behavior and content metrics are correlated with important events like elections. 
Our experimental results confirm that our user stratification strategy successfully brings out the unusual behavior of the extreme tweeters. The strategy itself is fairly generic and can be applied to any social media platform. In terms of results, we find that the highly-connected group we call ``anomalous'' indeed exhibits a ``clannish'' behavior because of their strong and narrowly-scoped within-group interactions that markedly differs from their across-group behavior. %Our study also reveals that the same group behaves differently with other high-volume tweeters, and regular users, 
The fact that their ``reach'' increases with time, especially leading up to the US mid-term elections, suggests that the group has strong political beliefs and forms a strong trust network of its own  that establishes contact with a high number of users but carefully controls content outside the group. Similarly,  the ``anomalous'' aspect of their behavior dramatically comes into play two weeks before the elections, suggesting a motivation to reinforce each others' beliefs and perhaps an intention to influence others in their mention network.

Our future work will elaborate this case study to uncover other characteristics of these user groups and their interactions. We will conduct a larger comparative study across more diverse time periods, a larger cross-section of users, and user behavior in different hashtag communities. We will also investigate more computationally efficient methods of finding interest narrowness, and user stratification. 
%We conduct experiments on a political tweet  data set and the time interval of interest is around the 2018 United States mid-term elections held on Nov 6. The results show some interesting phenomenons. When it is approaching  Nov 6, the number of extreme tweeters grows and they have an increasingly  narrower topic interest, in addition,  the anomalous users have  much more intense connections with each other and with other ETTs  than other time periods. After election, however, their connection suddenly become much looser and their topic interest also grows  broader.   

\bibliographystyle{abbrv}
\bibliography{paper}
\end{document}